\documentclass[10pt]{article}
\usepackage[textures]{graphics}

\parskip=\bigskipamount
\newcommand{\beq}{\begin{equation}}
\newcommand{\eeq}{\end{equation}}
\newcommand{\beqa}{\begin{eqnarray}}
\newcommand{\eeqa}{\end{eqnarray}}
\newcommand{\ket} [1] {\vert #1 \rangle}
\newcommand{\bra} [1] {\langle #1 \vert}

\newcommand{\proj}[1]{\ket{#1}\bra{#1}}

\newcommand{\mod}{~{\rm mod}~}

\begin{document}

\title{{\sc Cloning a Qutrit}
\thanks{To appear in the Journal of Modern Optics
for the special issue on ``Quantum Information: Theory, Experiment and
Perspectives''. Proceedings of the ESF Conference, Gdansk, July 10-18, 2001.}}

\author{Nicolas J. Cerf,$^{1,2}$ Thomas Durt,$^3$ and Nicolas Gisin$^4$\\
\\
$^1$ Ecole Polytechnique, CP 165, Universit\'e Libre de Bruxelles,\\
1050 Brussels, Belgium\\
$^2$ Jet Propulsion Laboratory, California Institute of Technology,\\
Pasadena, California 91109, USA\\
$^3$ Toegepaste Natuurkunde en Fotonica, Vrije Universiteit Brussel,\\
Pleinlaan 2, 1050 Brussels, Belgium\\
$^4$ Group of Applied Physics - Optique, Universit\'e de Gen\`eve,\\
20 rue de l'Ecole de M\'edecine, Gen\`eve 4, Switzerland}

\date{August 2001}

\maketitle

\begin{abstract}
We investigate several classes of state-dependent quantum cloners
for three-level systems. These cloners optimally duplicate
some of the four maximally-conjugate bases with an equal fidelity, thereby
extending the phase-covariant qubit cloner to qutrits. Three distinct classes
of qutrit cloners can be distinguished, depending on two, three, or four
maximally-conjugate bases are cloned as well (the latter 
case simply corresponds to the universal qutrit cloner). 
These results apply to symmetric as well as
asymmetric cloners, so that the balance between the fidelity of the two clones
can also be analyzed.
\end{abstract}

\section{Introduction}

Since its inception, quantum information theory has traditionally been
concerned
with informational processes involving two-level quantum systems, known as
qubits.
For example, quantum teleportation, quantum cryptography, quantum computation,
or quantum cloning were all developed using qubits as fundamental units of
quantum information\cite{preskill}. Over the last few years, however,
there has been a growing interest in quantum informational processes
based on multi-level or even continuous-spectrum systems.
There are several reasons for this. Firstly, higher-dimensional
quantum informational processes 
seem to be more efficient in certain situations. For example,
multi-level quantum cryptographic schemes can be shown to be more secure
against eavesdropping than their qubit-based counterparts\cite{bourennane}.
Second, the present experimental context makes it reasonable
to consider the manipulation of more-than-two-level quantum information
carriers. For example, the time-bin implementation of qubits can be relatively
straightforwardly extended to three of more time bins\cite{gisin}. Quantum
computation over continuous variables also seems to be a promising avenue,
as it can be carried out by manipulating squeezed states of light with
only linear optics elements\cite{braunstein}.

In this paper, we extend the concept of quantum cloning from qubits
to qutrits (quantum three-level systems). In spite of this apparently simple
incremental step of just one more dimension, the question of cloning
turns out to be already significantly more complex for qutrits, suggesting that
the state-dependent cloning of multi-level systems is a rich field.
Quantum cloning is a concept that was first introduced in a seminal paper
by Buzek and Hillery\cite{buzek}, 
where a universal (or state-independent) and symmetric
$1\to 2$ cloning transformation was introduced for qubits. This transformation
was later extended to higher dimensions by Buzek
and Hillery\cite{buzek98} and by Werner\cite{werner},
but for the special case of universal (state-independent) cloning. In contrast,
we will rather focus here on non-universal (or state-dependent) cloning.
Our starting point will be a general characterization of asymmetric and
state-dependent $1\to 2$ cloning transformations 
for $N$-level systems, as described in Refs.~\cite{cerf-acta,cerf-jmo}.
Let us first review this
formalism before analyzing the special case of $N=3$ in details.

Consider an arbitrary state $\ket{\psi}$ 
in a $N$-dimensional Hilbert space of which we wish to produce two
(approximate) clones. The class of cloning transformations
we will analyze is such that, if the input state is $\ket{\psi}$, 
then the two output clones (called $A$ and $B$) are produced
in a mixture of the states $\ket{\psi_{m,n}}=U_{m,n}\ket{\psi}$:
\beqa \label{rho}
\rho_A&=&\sum_{m,n=0}^{N-1} p_{m,n} \proj{\psi_{m,n}} \nonumber \\
\rho_B&=&\sum_{m,n=0}^{N-1} q_{m,n} \proj{\psi_{m,n}}
\eeqa
where the unitary operators
\beq
U_{m,n}=\sum_{k=0}^{N-1} e^{2\pi i (kn/N)} \ket{k+m \mod N}\bra{k}
\eeq
correspond to {\em error} operators: $U_{m,n}$ shifts the state by $m$ units
(modulo $N$) in the computational basis, and multiplies it by a phase
so as to shift its Fourier transform by $n$ units (modulo $N$). 
Of course, $U_{0,0}=I$, which corresponds to no error. In the special case
of a qubit ($N=2$), we have $U_{1,0}=\sigma_x$, $U_{0,1}=\sigma_z$,
and $U_{1,1}=-i \sigma_y$, and the corresponding class of so-called
Pauli cloners can be investigated exhaustively\cite{cerf-prl,cerf-nasa}.

From Eq.~(\ref{rho}), it is clear that the clones $A$ and $B$ are characterized
in general by the weight functions $p_{m,n}$ and $q_{m,n}$, respectively.
As we will see below, the class of cloners we will restrict our attention to
are defined by a particular relation between these weight
functions\cite{cerf-acta,cerf-jmo}. More specifically, we will focus
on cloners satisfying
\beqa \label{p_q_}
p_{m,n} &=& |a_{m,n}|^2 \nonumber \\
q_{m,n} &=& |b_{m,n}|^2
\eeqa
where $a_{m,n}$ and $b_{m,n}$ are two (complex) amplitude functions
that are dual under a Fourier transform:
\beq  \label{FT}
b_{m,n} = {1\over N} \sum_{x,y=0}^{N-1} e^{2\pi i (nx-my)/ N } a_{x,y}
\eeq
Of course, these amplitudes are normalized:
$\sum_{m,n} |a_{m,n}|^2 = \sum_{m,n} |b_{m,n}|^2 = 1$.
Interestingly enough, the cloners obeying Eqs. (\ref{p_q_}) and 
(\ref{FT}) form a fairly general class, which contains most cloners
discovered so far.

The Fourier transform that underlies the relation between the $p_{m,n}$'s
and the $q_{m,n}$'s is responsible for the  {\em complementarity}
between the quality of the two clones: if clone $A$ is very good
($a_{m,n}$ is a ``peaked'' function), 
then clone $B$ is very bad ($b_{m,n}$ is
a rather ``flat'' function, so that many error operators $U_{m,n}$
act on $\ket{\psi}$ with significant probabilities). Note that 
the relation between
these dual amplitude functions $a_{m,n}$ and $b_{m,n}$ can be reexpressed
in a simple way 
by associating them with Fourier transformed amplitude functions
$a^F_{m,n}$ and $b^F_{m,n}$ defined as
\beq \label{TR1}
c^F_{m,n} = {1\over \sqrt N}
\sum_{n'=0}^N e^{-2\pi i (nn'/N)} c_{m,n'}
\eeq
or, conversely,
\beq \label{TR3}
c_{m,n} = {1\over \sqrt N}
\sum_{n'=0}^N e^{2\pi i (nn'/N)} c^F_{m,n'}
\eeq
where $c_{m,n}$ stands for $a_{m,n}$ or $b_{m,n}$. We have
\beqa  \label{TR2}
b^F_{m,n} &=& {1\over \sqrt N}
\sum_{n'=0}^{N-1} e^{-2\pi i (nn'/N)} b_{m,n'} \nonumber\\
&=& {1\over \sqrt N}
\sum_{n'=0}^{N-1} e^{-2\pi i (nn'/N)}
{1\over N}\sum_{x,y=0}^{N-1} e^{2\pi i (n'x-my)/N}a_{x,y} \nonumber \\
&=& {1\over \sqrt N}
\sum_{y=0}^{N-1} e^{-2\pi i (my/N)} a_{n,y} \nonumber \\
&=& a^F_{n,m}
\eeqa
Therefore, the Fourier transformed amplitudes $a^F_{m,n}$ and $b^F_{m,n}$
of the two clones are simply {\em transposed} of each other,
which will be helpful
in the following. The balance between the quality of clones $A$ and $B$ can
be alternatively expressed by an entropic no-cloning
uncertainty relation that relates the probability distributions
$p_{m,n}$ and $q_{m,n}$ \cite{cerf-jmo}:
\beq
H[p_{m,n}]+H[q_{m,n}] \ge \log_2(N^2)
\eeq
where $H[p]$ denote the Shannon entropy of the probability distribution $p$.
This inequality is actually a special case of an information-theoretic
no-cloning uncertainty relation involving the {\em losses} 
of the channels yielding the two clones\cite{cerf-nasa}.
Also, more refined uncertainty relations can be found that express
the fact that the index $m$ of output $A$ is dual to the index $n$
of output $B$, and conversely\cite{cerf-jmo}.

Let us now describe the class of cloning transformations that
actually produce the clones characterized by Eq.~(\ref{rho}). For this,
we need first to define the set of $N^2$ generalized Bell states
for a pair of N-dimensional systems:
\beq
\ket{B_{m,n}}=N^{-1/2} \sum_{k=0}^{N-1} e^{2\pi i (kn/N)} \ket{k}\ket{k+m}
\eeq
with $m$ and $n$ ($0\le m,n \le N-1$) labelling these Bell states.
We will characterize the cloning transformation by
assuming that the cloner input is prepared in the joint state 
$\ket{B_{0,0}}$ together with a ($N$-dimensional) system called 
a {\em reference} system and denoted as $R$. 
We will consider a unitary cloning transformation 
${\cal U}_{\rm cl}$ acting on this input system together
with two additional $N$-dimensional systems prepared each in an
initial state $\ket{0}$: a blank copy  
and the cloning machine itself. After transformation,
the input system and the blank copy become respectively the clones
$A$ and $B$, while the cloning machine denoted as $C$
can be traced over. We will thus be interested 
in the joint state (after cloning) 
of the reference $R$, the two clones ($A$ and $B$),
and the cloning machine $C$, that is
\beq
(I_R \otimes {\cal U}_{\rm cl}) \; \ket{B_{0,0}} \ket{0} \ket{0}
=\ket{\Psi}_{RABC}
\eeq
More specifically, we will only consider joint states which can be written as
\beq
\ket{\Psi}_{RABC}
= \sum_{m,n=0}^{N-1} a_{m,n} \; \ket{B_{m,n}}_{R,A} \ket{B_{m,-n}}_{B,C}
= \sum_{m,n=0}^{N-1} b_{m,n} \; \ket{B_{m,n}}_{R,B} \ket{B_{m,-n}}_{A,C}
\eeq
with $a_{m,n}$ and $b_{m,n}$ being related by Eq. (\ref{FT}).
This construction is very useful because one can easily express
the output state resulting from cloning an arbitrary input state $\ket{\psi}$
simply by projecting the reference system onto an appropriate state.
Indeed, before cloning, projecting $R$ onto state $\ket{\psi^*}$
amounts to project the input system onto $\ket{\psi}$ since these two systems
are in state $\ket{B_{0,0}}$. Therefore, as this projection 
of $R$ onto $\ket{\psi^*}$ can as well be performed {\em after} cloning,
it is easy to write the resulting joint state of the two clones 
and the cloning machine when the input state is $\ket{\psi}$. 
Using $\ket{B_{m,n}}=(I\otimes U_{m,n})\ket{B_{0,0}}$, we get
\beq  \label{transfo}
\ket{\psi}  \to
\sum_{m,n=0}^{N-1} a_{m,n} \; U_{m,n}\ket{\psi}_A \ket{B_{m,-n}}_{B,C}
= \sum_{m,n=0}^{N-1} b_{m,n} \; U_{m,n}\ket{\psi}_B \ket{B_{m,-n}}_{A,C}
\eeq
Now, it is easy to check that tracing over systems $B$ and $C$ (or $A$ and $C$)
yields the expected final states of clone $A$ (or clone $B$), in accordance
with Eq.~(\ref{rho}). Thus, the $N^2$ amplitudes $a_{m,n}$ (or $b_{m,n}$)
completely define the state after cloning, Eq.~(\ref{transfo}), so
they completely characterize the class of cloning transformations
of interest here. 

Finally, let us see how the cloning fidelity can be calculated
based on these amplitude matrices $a_{m,n}$ or $b_{m,n}$.
The fidelity of the first clone when copying a state $\ket{\psi}$
can be written, in general, as
\beq \label{fidelity}
F_A=\langle\psi|\rho_A|\psi\rangle
= \sum_{m,n=0}^{N-1} |a_{m,n}|^2  |\langle\psi|\psi_{m,n}\rangle|^2
\eeq
(Of course, the same relation can be used for the second clone by replacing
$a_{m,n}$ by $b_{m,n}$.) For example, for any state $\ket{k}$ ($k=0,\ldots
N-1$)
in the computation basis, the fidelity of the first clone is equal to
\beq   \label{F_A}
F_A=\sum_{n=0}^{N-1} |a_{0,n}|^2
\eeq
As we will see later on for $N=3$, the cloning fidelity for other bases
can also be written as a sum of three squared terms of the $a_{m,n}$ matrix.
This will make it possible to express constraints on the state-dependent
cloners of interest.

In the rest of this paper, we will use this general characterization of cloning
in order to investigate the state-dependent cloning of a qutrit.
Four maximally-conjugate bases can be defined in a three-dimensional space:
these bases are such that any basis state in one basis
has equal squared amplitudes when expressed in any other basis.
We will analyze transformations that optimally clone a subset of these
four maximally-conjugate bases for a qutrit. Three interesting situations occur
depending on whether we consider a subset of two or three of these bases,
or all four bases. The case of a qubit is also treated in Appendix~\ref{qubit}
for completeness\footnote{The reader may want to
read this Appendix first as an introduction to the concept of state-dependent
cloning. The case of a qubit is indeed much simpler to treat than that of a
qutrit.}.

\section{Cloning a three-level system}

In a three-dimensional Hilbert space, one can define four maximally-conjugate
(or mutually unbiased) bases\cite{bechmann}.
Conventionally, one chooses the first basis to be simply the computation basis
$\{ \ket{0},\ket{1},\ket{2} \}$. The second basis is defined as
\beqa
\ket{0'}&=&{1\over\sqrt{3}}(\ket{0}+\ket{1}+\ket{2}), \nonumber\\
\ket{1'}&=&{1\over\sqrt{3}}(\ket{0}+\gamma\ket{1}+\gamma^2\ket{2}), \nonumber\\
\ket{2'}&=&{1\over\sqrt{3}}(\ket{0}+\gamma^2\ket{1}+\gamma\ket{2})
\eeqa
where $\gamma=e^{2\pi i /3}$. Similarly, the third basis is defined as
\beqa
\ket{0''}&=&{1\over\sqrt{3}}(\ket{0}+\ket{1}+\gamma\ket{2}), \nonumber\\
\ket{1''}&=&{1\over\sqrt{3}}(\ket{0}+\gamma\ket{1}+\ket{2}), \nonumber\\
\ket{2''}&=&{1\over\sqrt{3}}(\gamma\ket{0}+\ket{1}+\ket{2})
\eeqa
while the fourth basis is defined as
\beqa
\ket{0'''}&=&{1\over\sqrt{3}}(\ket{0}+\ket{1}+\gamma^2\ket{2}), \nonumber\\
\ket{1'''}&=&{1\over\sqrt{3}}(\ket{0}+\gamma^2\ket{1}+\ket{2}), \nonumber\\
\ket{2'''}&=&{1\over\sqrt{3}}(\gamma^2\ket{0}+\ket{1}+\ket{2})
\eeqa
It is easy to check that the scalar product between any two basis states
belonging to two distinct bases is $1/\sqrt{3}$, as expected. We can also
check that the first and second bases are connected by a discrete Fourier
transform. The same relation holds for the third and fourth bases.

Let us start by calculating the action of the nine error
operators $U_{m,n}$ on these basis states. Within each of these four bases,
it can be shown that applying $U_{m,n}$ to one basis state $\ket{k}$
yields either $\ket{k}$ or $\ket{k+1 \mod 3}$ or $\ket{k+2 \mod 3}$
up to a phase $\gamma$ or $\gamma^2$. Using this property, we can express
the fidelity of the first (or second) clone in each basis. The fidelity
of one of the clones when copying a state $\ket{\psi}$ is defined as
\beq
F=\langle\psi|\rho|\psi\rangle
\eeq
where $\rho$ is defined in Eq.~(\ref{rho}). Note that,
unlike the situation for a qubit, there are {\em two} possible errors when
copying the basis state $\ket{k}$ (in a given basis) for a qutrit
depending on it being transformed into $\ket{k+1 \mod 3}$ or 
$\ket{k+2 \mod 3}$.
Therefore, we define {\em two} disturbances $D_1$ and $D_2$ corresponding
to these two errors. Remembering that the state of the first clone is
completely
characterized by the matrix
\beq
(p_{m,n})=\left(
\begin{array}{lll}
p_{0,0} & p_{0,1} & p_{0,2}\\
p_{1,0} & p_{1,1} & p_{1,2}\\
p_{2,0} & p_{2,1} & p_{2,2}
\end{array}
\right)
\eeq
we can calculate the fidelity and the two disturbances when cloning
any basis state in any basis. For example,
for the first maximally-conjugate basis, we have
\beqa
F  &=&p_{0,0}+p_{0,1}+p_{0,2} \\
D_1&=&p_{1,0}+p_{1,1}+p_{1,2} \\
D_2&=&p_{2,0}+p_{2,1}+p_{2,2}
\eeqa
The cloning of the three last maximally-conjugate bases can be treated
together by considering the state
\beq
\ket{\psi_{0}} = {1\over\sqrt 3}
\left(\ket{0}+e^{i\alpha}\ket{1}+e^{i\beta}\ket{2}\right)
\eeq
with arbitrary $\alpha$ and $\beta$.
By direct computation, we get the fidelity
\beqa  \label{F1}
\lefteqn{  F=\bra{\psi_{0}} \rho \ket{\psi_{0}} = p_{0,0} + {1\over 3}
(p_{1,0}+p_{2,0}+p_{1,2}+p_{2,1}+p_{1,1}+p_{2,2})
} \nonumber \\
&+& {2\over 9}
(p_{1,0}+p_{2,0}) \Big[
\cos(\alpha+\beta)+\cos(\alpha-2\beta)+\cos(\beta-2\alpha)\Big]
   \nonumber \\
&+& {2\over 9}
(p_{1,2}+p_{2,1}) \Big[ \cos(\alpha+\beta+2\pi/3) +
\cos(\alpha-2\beta+2\pi/3) + \cos(\beta-2\alpha+2\pi/3) \Big]
  \nonumber \\
&+& {2\over 9}
(p_{1,1}+p_{2,2}) \Big[ \cos(\alpha+\beta-2\pi/3) +
\cos(\alpha-2\beta-2\pi/3) + \cos(\beta-2\alpha-2\pi/3) \Big]
\eeqa
Before calculating the disturbances, we need first to define the states
\beqa
\ket{\psi_{1}}&=&{1\over \sqrt 3}\left(\ket{0}+\gamma e^{i\alpha}\ket{1}
+\gamma^2 e^{i\beta} \ket{2}\right) \nonumber \\
\ket{\psi_{2}}&=&{1\over \sqrt 3}\left(\ket{0}+\gamma^2 e^{i\alpha}\ket{1}
+\gamma e^{i\beta}\ket{2}\right)
\eeqa
which, together with $\ket{\psi_0}$, form an orthonormal basis.
We can easily rewrite the second, third, and fourth maximally-conjugate
bases in the
form $\{\ket{\psi_{0}},\ket{\psi_{1}},\ket{\psi_{2}}\}$
for well chosen values of $\alpha$ and $\beta$. This
is of course not true for the first (computational) basis.
We can now calculate the disturbances,
\beqa  \label{D1}
\lefteqn{ D_1 =\bra{\psi_{1}} \rho \ket{\psi_{1}} = p_{0,1} + {1\over 3}
(p_{1,1}+p_{2,1}+p_{1,0}+p_{2,2}+p_{1,2}+p_{2,0})
} \nonumber \\
&+& {2\over 9}
(p_{1,1}+p_{2,1}) \Big[
\cos(\alpha+\beta)+\cos(\alpha-2\beta)+\cos(\beta-2\alpha)\Big]
   \nonumber \\
&+& {2\over 9}
(p_{1,0}+p_{2,2}) \Big[ \cos(\alpha+\beta+2\pi/3) +
\cos(\alpha-2\beta+2\pi/3) + \cos(\beta-2\alpha+2\pi/3) \Big]
  \nonumber \\
&+& {2\over 9}
(p_{1,2}+p_{2,0}) \Big[ \cos(\alpha+\beta-2\pi/3) +
\cos(\alpha-2\beta-2\pi/3) + \cos(\beta-2\alpha-2\pi/3) \Big]
\eeqa
and
\beqa  \label{D2}
\lefteqn{ D_2 =\bra{\psi_{2}} \rho \ket{\psi_{2}} = p_{0,2} + {1\over 3}
(p_{1,2}+p_{2,2}+p_{1,1}+p_{2,0}+p_{1,0}+p_{2,1})
} \nonumber \\
&+& {2\over 9}
(p_{1,2}+p_{2,2}) \Big[
\cos(\alpha+\beta)+\cos(\alpha-2\beta)+\cos(\beta-2\alpha)\Big]
   \nonumber \\
&+& {2\over 9}
(p_{1,1}+p_{2,0}) \Big[ \cos(\alpha+\beta+2\pi/3) +
\cos(\alpha-2\beta+2\pi/3) + \cos(\beta-2\alpha+2\pi/3) \Big]
  \nonumber \\
&+& {2\over 9}
(p_{1,0}+p_{2,1}) \Big[ \cos(\alpha+\beta-2\pi/3) +
\cos(\alpha-2\beta-2\pi/3) + \cos(\beta-2\alpha-2\pi/3) \Big]
\eeqa
It can be easily checked that $F$, $D_1$, and $D_2$ are invariant if we replace
$(\alpha,\beta)$ by $(\alpha\pm 2\pi/3,\beta\pm 4\pi/3)$, and that these
phase-shifts simply permute cyclically the states $\ket{\psi_{0}}$,
$\ket{\psi_{1}}$, and $\ket{\psi_{2}}$.
Therefore, the values of $F$, $D_1$ and $D_2$ are invariant
under a cyclic permutation of the states of the maximally-conjugate bases.
Note that all the states of the second maximally-conjugate basis fulfill
$\alpha+\beta=\alpha-2\beta=\beta-2\alpha=0$.
Similarly, in the third and fourth maximally-conjugate bases, we have
$\alpha+\beta=\alpha-2\beta=\beta-2\alpha=2\pi/3$ and
$-2\pi/3$, respectively. For those states, Eqs.~(\ref{F1}),
(\ref{D1}), and (\ref{D2}) can be quite simplified. 
For instance, the fidelity and
disturbances when cloning any basis state of the second
maximally-conjugate basis are given by
\beqa
F'&=&p_{0,0}+p_{1,0}+p_{2,0}  \nonumber\\
D_1'&=&p_{0,1}+p_{1,1}+p_{2,1}  \nonumber\\
D_2'&=&p_{0,2}+p_{1,2}+p_{2,2}
\eeqa
For the third basis, we have
\beqa
F''&=&p_{0,0}+p_{1,1}+p_{2,2}  \nonumber\\
D_1''&=&p_{0,1}+p_{1,2}+p_{2,0}  \nonumber\\
D_2''&=&p_{0,2}+p_{1,0}+p_{2,1}
\eeqa
while the fourth basis yields
\beqa
F'''&=&p_{0,0}+p_{1,2}+p_{2,1}  \nonumber\\
D_1'''&=&p_{0,1}+p_{1,0}+p_{2,2}  \nonumber\\
D_2'''&=&p_{0,2}+p_{1,1}+p_{2,0}
\eeqa

In the following, we will be interested in extending to a 3-dimensional space
the so-called phase-covariant qubit cloner described in the Appendix.
Two extensions can be considered,
depending on two or three of the maximally-conjugate bases are copied
equally well. The cloner that copies all the four bases with an equal fidelity
is simply the universal cloner, as discussed in the last Section.

\section{Optimal cloner of two maximally-conjugate bases}

Here, we consider a state-dependent cloner that clones equally well the
third and fourth maxi\-mally-conjugate bases.
This imposes that
\beqa
p_{1,1}+p_{2,2}=p_{1,2}+p_{2,1}\nonumber\\
p_{1,2}+p_{2,0}=p_{1,0}+p_{2,2}\nonumber\\
p_{1,0}+p_{2,1}=p_{1,1}+p_{2,0}
\eeqa
It is easy to deduce from these constraints together with Eq.~(\ref{F1}) that
the cloning fidelity for an arbitrary state $\ket{\psi_0}$
is equal to
\begin{eqnarray}
\lefteqn{
F=p_{0,0}+{1\over 3}(p_{1,0}+p_{2,0}+2(p_{1,2}+p_{2,1}))
} \nonumber \\
&+& {2\over 9}\Big[ p_{1,0}+p_{2,0}-(p_{1,2}+p_{2,1})\Big] \;
\Big[ \cos(\alpha+\beta)+\cos(\alpha-2\beta)+\cos(\beta-2\alpha)\Big]
\end{eqnarray}
The function
$\cos(\alpha+\beta)+\cos(\alpha-2\beta)+\cos(\beta-2\alpha)$
reaches its extremal value $-3/2$ when
$\alpha+\beta=\alpha-2\beta=\beta-2\alpha=2\pi/ 3$
or $-2\pi/ 3$, that is, when $\ket{\psi_{0}}$ belongs to the third
or fourth basis. Therefore, when the second basis is not cloned as well as the
third and fourth maximally-conjugate bases, i.e. when $p_{1,0}+p_{2,0} <
p_{1,1}+p_{2,2}$,
there exists no state of the form ${1\over \sqrt 3}
(\ket{0}+e^{i\alpha}\ket{1}+e^{i\beta}\ket{2})$ outside the third and
fourth bases that is equally well cloned.
Similarly, we expect that when the first and
second bases are not cloned as well as the third and fourth
maximally-conjugate bases, i.e. when
$p_{0,1}+p_{0,2} < p_{1,1}+p_{2,2}$ and $p_{1,0}+p_{2,0} < p_{1,1}+p_{2,2}$,
then there exists no state at all outside
the third and fourth bases that gets equally well cloned.

Let us now consider a state-dependent qutrit cloner that is characterized
by the
amplitude matrix
\begin{equation}
(a_{m,n})= \left(\begin{array}{ccc}
v & y & y \\
y & x & x \\
y & x & x \\
\end{array}\right)
\end{equation}
where $v,x,$ and $y$ are real parameters obeying
the normalization condition $v^2+4x^2+4y^2=1$. This matrix
corresponds to the probability matrix $p_{m,n}=a_{m,n}^2$.
It is easy to check that this cloner results in
a same fidelity (and same disturbances: $D_1=D_2$) for all the basis states of
the two last bases
$\{\ket{0''},\ket{1''},\ket{2''}\}$ and
$\{\ket{0'''},\ket{1'''},\ket{2'''}\}$:
\beqa
F''&=& F''' \;=\; v^2+2x^2  \nonumber \\
D_{1,2}''&=& D_{1,2}''' \;=\; x^2+2y^2
\eeqa
Of course, we have  $F+D_1+D_2=1$.
Using Eqs.~(\ref{TR1}), (\ref{TR3}), and (\ref{TR2}),
we get
\begin{equation}
(a^F_{m,n})={1\over \sqrt 3}
\left(\begin{array}{ccc}  v+2y & v-y & v-y \\
y+2x & y-x & y-x\\
y+2x & y-x & y-x \\
\end{array}\right)
\end{equation}
\begin{equation}
(b^{F}_{m,n})= {1\over \sqrt 3}
\left(\begin{array}{ccc}  v+2y &y+2x  & y+2x \\
v-y & y-x & y-x\\
v-y & y-x & y-x \\
\end{array}\right)
\end{equation}
\begin{equation}
(b_{m,n})= {1\over 3}
\left(\begin{array}{ccc}
v + 4x+4y&v-2x+y &v-2x+y  \\
v-2x+y & v+x -2y &v+x -2y \\
v-2x+y &v+x -2y   & v+x -2y \\
\end{array}\right)
\end{equation}
so that the matrix $b_{m,n}$ characterizing the second clone has the same
form as
$a_{m,n}$ with the substitution:
\beqa
v &\to& (v+4x+4y)/3 \\
x &\to& (v+x-2y)/3 \\
y &\to& (v-2x+y)/3
\eeqa
Consequently, the states of the two last bases are again copied with
the same fidelity (and same disturbances) onto the second clone:
\beqa
\tilde F &=& (v^2+6x^2+8y^2+4vx+8xy)/3 \\
\tilde D_{1,2} &=& (v^2+3x^2+2y^2-2vx-4xy)/3
\eeqa
We will be now interested in finding the optimal cloner, that is the cloner
that maximizes the fidelity of the second clone
for a given fidelity of the first clone. Maximizing $\tilde F$
with the constraint that $F$ is given and using the normalization condition
yields the solution
\beqa
v&=&F\\
x&=&\sqrt{F(1-F)/2}\\
y&=&(1-F)/2
\eeqa
Hence, the fidelity of the second clone can be written as
a function of the fidelity of the first clone
\beq  \label{2-b-1}
\tilde F={2-F\over 3} + {2\sqrt{2}\over 3}\sqrt{F(1-F)}
\eeq
which expresses the complementarity between the clones.
As expected, $F=1$ implies $\tilde F=1/3$, and conversely.
An interesting special case is the symmetric cloner, which
yields two clones of equal fidelity
\beq  \label{2-b-2}
F=\tilde F={1\over 2}+{1\over\sqrt{12}} \simeq 0.789
\eeq
It should be noted that Eqs.~(\ref{2-b-1}) and (\ref{2-b-2})
hold regardless which bases are optimally cloned, provided
that two of them are equally cloned. The two remaining ones
are then copied with a lower fidelity $v^2+2y^2={1\over 2}+{1\over 2\sqrt{12}}
\simeq 0.644$. Note also that a more general cloner could be constructed
for which these two remaining bases are not cloned with an equal fidelity,
but it will not be considered here.

\section{Optimal cloner of three maximally-conjugate bases}

Now, we consider a state-dependent cloner that clones equally well the
three last maximally-conjugate bases and for which $D_1=D_2$. Again,
our result will actually be independent of which three bases are
optimally cloned, so we only consider the last three ones for
simplicity. This imposes that
\beqa  \label{cons}p_{0,1}+p_{1,1}+p_{2,1}  = p_{0,2}+p_{1,2}+p_{2,2}
\nonumber\\
p_{1,0}+p_{2,0}=p_{1,1}+p_{2,2}=p_{1,2}+p_{2,1}\nonumber\\
p_{1,1}+p_{2,1}=p_{1,2}+p_{2,0}=p_{1,0}+p_{2,2}\nonumber\\
p_{1,2}+p_{2,2}=p_{1,0}+p_{2,1}=p_{1,1}+p_{2,0}
\eeqa
It is easy to deduce from these constraints together with Eqs.~(\ref{F1}),
(\ref{D1}), and (\ref{D2}) that the cloning fidelity for an arbitrary state
$\ket{\psi_{0}}={1\over \sqrt 3}(\ket{0}+e^{i\alpha}\ket{1}+e^{i\beta}\ket{2})$
is simply given by
\begin{equation}
F=p_{0,0}+p_{1,0}+p_{2,0}
\end{equation}
that is, it coincides with the cloning fidelity of the elements of
the three last bases. This simplification occurs because of cyclical
compensations in Eq.~(\ref{F1}) which
originate from the fact that the number of bases that are equally
well cloned here is equal to the dimension of the Hilbert space (3 in the
present case). This situation generalizes the one encountered with
the phase-covariant cloner for a qubit (see Appendix~\ref{qubit}). In
that case, the cloner that clones
equally well two maximally-conjugate bases in a Hilbert space of dimension 2
can be shown to clone equally well all the states of an equator
of the Bloch sphere. In the present case, the cloner
that clones equally well two plus one maximally-conjugate bases
clones equally well the generalized equator, i.e.,
a 1+1 dimensional variety that contains all the states of the form
$\ket{\psi_{0}}={1\over \sqrt
3}(\ket{0}+e^{i\alpha}\ket{1}+e^{i\beta}\ket{2})$.

It can be shown that the general solution of Eq.~(\ref{cons}) is
a probability matrix $p_{m,n}$ of the form
\begin{equation}  \label{3bcloner}
(p_{m,n})= \left(\begin{array}{ccc}
v^2 & x^2 & x^2 \\
y^2 & y^2 & y^2 \\
z^2 & z^2  & z^2 \\
\end{array}\right)
\end{equation}
For instance, we have that $p_{1,0}+p_{2,0}-(p_{1,1}+p_{2,1})$ =
$(p_{1,1}+p_{2,2}+p_{1,2}+p_{2,1})/2
-(p_{1,2}+p_{2,0}+p_{1,0}+p_{2,2})/2$
= $(p_{1,1}+p_{2,1}-(p_{2,0}+p_{1,0}))/2$, so that
$p_{1,0}+p_{2,0}=p_{1,1}+p_{2,1}$. But we have
$p_{0,0}+p_{1,0}+p_{2,0}=p_{0,0}+p_{1,1}+p_{2,2}$, so that
$p_{2,1}=p_{2,2}$.
We deduce in a similar way that one must have
$p_{2,0}=p_{2,1}=p_{2,2}$,
$p_{1,0}=p_{1,1}=p_{1,2}$, and $p_{0,1}=p_{0,2}.$
It is easy to check that these conditions are also
sufficient conditions.

\subsection{Symmetric cloner}

Let us now consider the symmetric state-dependent cloner that clones
equally well the three last bases 
and is characterized by the amplitude matrix

\begin{equation}(a_{m,n})= \left(\begin{array}{ccc}
x+y+z & x+\alpha y+\alpha^2 z & x+\alpha^2 y+\alpha z \\
y & y & y \\
z & z  & z \\
\end{array}\right)\end{equation} where $x,y,$ and $z$ are real parameters
and with the
normalisation condition $3x^2+6x^2+6z^2=1$. This matrix
corresponds to the
probability matrix $p_{m,n}=a_{m,n}^2$. It is easy to check that this
cloner results in
a same fidelity (and same disturbance) for all basis states of the three
last bases
$\{\ket{0'},\ket{1'},\ket{2'}\}$, $\{\ket{0''},\ket{1''},\ket{2''}\}$ and
$\{\ket{0'''},\ket{1'''},\ket{2'''}\}$:
\begin{equation}
F'=F''=F'''=x^2+2y^2+2z^2+2xy+2yz+2xz
\end{equation}
\begin{equation}
D_{1,2}'=D_{1,2}''=D_{1,2}'''=x^2+2y^2+2z^2-xy-yz-xz
\end{equation}
Of course we have $F+D_1+D_2=1.$
By Eqs.~(\ref{TR1}), ~(\ref{TR2}), ~(\ref{TR3}):
\begin{equation}
(a^F_{m,n})={1\over \sqrt 3}
\left(\begin{array}{ccc}  3x & 3y & 3z \\
3y & 0 & 0 \\
3z & 0 & 0 \\
\end{array}\right)
\end{equation}
\beq 
(b^{F}_{m,n})= (a^F_{m,n})
\eeq 
\beq
(b_{m,n})=(a_{m,n})
\eeq 
which shows that this cloner is symmetric. The cloner is
optimal when the
fidelity $x^2+2y^2+2z^2+2xy+2yz+2xz$ is maximal under
the constraint
that $x^2+2y^2+2z^2=1/3$. By the method of Lagrange, we obtain that the
fidelity is extremal when the following equations are satisfied:
\begin{eqnarray}
y+z=\lambda x\nonumber \\
x+z=2\lambda y\nonumber \\
x+y=2\lambda z
\end{eqnarray} 
where $\lambda$ is a Lagrange multiplier. From
the last two equations, we deduce that either $\lambda$ = $-{1\over 2}$
or $y=z$. If
$\lambda$ = $-{1\over 2}$, then $x=0$, $y=-z$ and $F={1\over 6}$
which is a minimum.
If $y=z$, then $\lambda={1\pm\sqrt{17}\over 4}$ and
$F={5\pm\sqrt{17}\over 12}$. The
maximal fidelity is thus equal to
\beq
F_{max}={5+\sqrt{17}\over 12} \simeq 0.760
\eeq
It corresponds to an amplitude matrix
\beq
(a_{m,n})= \left(\begin{array}{ccc}
x+2y & x-y & x-y \\
y & y & y \\
y & y & y \\
\end{array}\right)
\eeq
with $x=\sqrt{{17-\sqrt{17}\over 102}}$ and
$y=\sqrt{{17+\sqrt{17}\over 408}}.$
It should be noted that this symmetric cloner exactly coincides with the
so-called double-phase covariant qutrit cloner that was independently
derived in Ref.~\cite{presti}.

\subsection{Asymmetric cloner}

Let us now consider the asymmetric state-dependent cloner that clones
equally well the three last maximally-conjugate bases,
and is characterized by the amplitude matrix
\begin{equation}\label{asym}
(a_{m,n})= \left(\begin{array}{ccc}
v & y & y \\
x & x & x \\
x & x  & x \\
\end{array}\right)
\end{equation}
where $v,x,$ and $y$ are real parameters and with the
normalisation condition $v^2+6x^2+2y^2=1$. This matrix
corresponds to the probability matrix $p_{m,n}=a_{m,n}^2$. It is easy
to check that this cloner results in
a same fidelity (and same disturbances) for all basis states of the three
last bases $\{\ket{0'},\ket{1'},\ket{2'}\}$,
$\{\ket{0''},\ket{1''},\ket{2''}\}$ and
$\{\ket{0'''},\ket{1'''},\ket{2'''}\}$:
\begin{equation}F'=F''=F'''=v^2+2x^2\end{equation}
\begin{equation}D_{1,2}'=D_{1,2}''=D_{1,2}'''=2x^2+y^2\end{equation}
Of course, we have again $F+D_1+D_2=1.$ By use of Eqs.~(\ref{TR1}),
(\ref{TR3}), and (\ref{TR2}), we get
\begin{equation}(a^F_{m,n})={1\over \sqrt 3}
\left(\begin{array}{ccc}  v+2y & v-y & v-y \\
3x & 0 & 0 \\
3x & 0 & 0 \\
\end{array}\right)\end{equation}
\begin{equation}(b^{F}_{m,n})={1\over \sqrt 3}
\left(\begin{array}{ccc}  v+2y & 3x & 3x \\
v-y & 0 & 0 \\
v-y & 0 & 0 \\
\end{array}\right)\end{equation}
\begin{equation}(b_{m,n})={1\over  3}
\left(\begin{array}{ccc}  v+6x+2y & v-3x+2y & v-3x+2y \\
v-y & v-y & v-y \\
v-y & v-y & v-y \\
\end{array}\right)
\end{equation}
Hence, for the second clone, the matrix $b_{m,n}$ has again
the same form as $a_{m,n}$ with the substitution
\beqa
v &\to& (v+6x+2y)/3 \\
x &\to& (v-y)/3 \\
y &\to& (v-3x+2y)/3
\eeqa
so that the states of the last three bases are all copied 
onto the second clone with a same fidelity (and same disturbances):
\beqa
\tilde F &=& (v^2+12x^2+2y^2+4vx+8xy)/3 \\
\tilde D_{1,2} &=& (v^2+3x^2+2y^2-2vx-4xy)/3
\eeqa
For the optimal cloner, we need to maximize $\tilde F$
for a given value of $F$ using the normalisation condition, just as
before. However, in contrast with the case of the asymmetric cloner for two maximally-conjugate bases, we have found no simple analytical solution
for this problem. A numerical solution and its connections with
quantum cryptography will be discussed elsewhere. Note that an asymmetric
state-dependent cloner could be constructed for which the last three bases
are all copied equally well but with a more general matrix $a_{m,n}$
than in Eq.~(\ref{asym}). It can be shown
however, that the {\em optimal} such cloner must necessarily obey
Eq.~(\ref{asym}) so that this possibility will not be considered here.

\section{Optimal cloner of all the maximally-conjugate bases}


Let us finally consider an asymmetric cloner that copies equally well all four
maximally-conjugate bases and for which $D_1=D_2$. We already showed
that the constraints (\ref{cons}) must be obeyed
for cloning equally well the last three maximally-conjugate
bases. In order to clone the fourth basis equally well, we must also impose
the additional constraints:
\beqa
p_{0,1}+p_{0,2}=p_{1,0}+p_{2,0}\nonumber\\
p_{1,0}+p_{1,2}=p_{0,1}+p_{2,1}\nonumber\\
p_{2,0}+p_{2,1}=p_{0,2}+p_{1,2}
\eeqa
Equivalently, using Eq.~(\ref{3bcloner}), we have
\begin{eqnarray}
2x^2=y^2+z^2 \nonumber \\
2y^2=x^2+z^2\nonumber\\
2z^2=x^2+y^2
\end{eqnarray}
Hence, $x^2=y^2=z^2$, and $p_{m,n}$ must be of the form
\begin{equation}
(p_{m,n})= \left(\begin{array}{ccc}
v^2 & x^2 & x^2 \\
x^2 & x^2 & x^2 \\
x^2 & x^2  & x^2 \\
\end{array}\right)
\end{equation}
It is thus natural to consider the following amplitude matrix
\begin{equation}
(a_{m,n})= \left(\begin{array}{ccc}
v & x & x \\
x & x & x \\
x & x  & x \\
\end{array}\right)
\end{equation}
where $v$ and $x$ are real parameters that satisfy
the normalization condition $v^2+8x^2=1$.
This matrix corresponds to the
probability matrix $p_{m,n}=a_{m,n}^2$.
By use of Eqs.~(\ref{TR1}), (\ref{TR3}), and (\ref{TR2}), we have
\begin{equation}(a^F_{m,n})={1\over \sqrt 3}
\left(\begin{array}{ccc}  v+2x & v-x & v-x \\
3x & 0 & 0 \\
3x & 0 & 0 \\
\end{array}\right)\end{equation}
\begin{equation}(b^{F}_{m,n})={1\over \sqrt 3}
\left(\begin{array}{ccc}  v+2x & 3x & 3x \\
v-x & 0 & 0 \\
v-x & 0 & 0 \\
\end{array}\right)\end{equation}
\begin{equation}(b_{m,n})={1\over 3}
\left(\begin{array}{ccc}  v+8x & v-x & v-x \\
v-x & v-x & v-x \\
v-x & v-x & v-x \\
\end{array}\right)\end{equation}
so that, for the second clone, the matrix
$b_{m,n}$ has the same form as $a_{m,n}$ with the
substitution:
\beqa
v &\to& {v + 8x\over 3} \\
x &\to& {v - x\over 3}
\eeqa
It is convenient here to change the variables $v$ and $x$ into
 $\alpha$ and $\beta$ according to
\beqa
v&=&\alpha+{\beta \over 3}\nonumber\\
x&=&{\beta \over 3}
\eeqa
so that we have
\begin{equation}
(a_{m,n})= \left(\begin{array}{ccc}
\alpha+{\beta\over 3} & {\beta\over 3} & {\beta\over 3} \\
{\beta\over 3}& {\beta\over 3} & {\beta\over 3} \\
{\beta\over 3} & {\beta\over 3}  & {\beta\over 3} \\
\end{array}\right)
\end{equation}
\begin{equation}
(b_{m,n})= \left(\begin{array}{ccc}
\beta+{\alpha\over 3} & {\alpha\over 3} & {\alpha\over 3} \\
{\alpha\over 3}& {\alpha\over 3} & {\alpha\over 3} \\
{\alpha\over 3} & {\alpha\over 3}  & {\alpha\over 3} \\
\end{array}\right)
\end{equation}
It is easy to check that this cloner results in
a same fidelity (and same disturbance) for any qutrit state:
\begin{equation}
F=\alpha^2+2{\alpha\beta\over 3}+{\beta^2\over
3}
\end{equation}
\begin{equation}
D_{1,2}={\beta^2\over 3}
\end{equation}
Of course we have $F+D_1+D_2=1.$
This is the special case of a state-independent (or universal) N-dimensional
cloner\cite{cerf-acta,cerf-jmo}, which can be obtained simply by letting
\beqa
a_{m,n}&=&\alpha \; \delta_{m,0} \delta_{n,0} + \beta /N \\
b_{m,n}&=&\beta \; \delta_{m,0} \delta_{n,0} + \alpha /N
\eeqa
This is consistent with Eq.~(\ref{FT})
since the constant function $1/N$ is the Fourier transform
of $\delta_{m,0} \, \delta_{n,0}$. Thus, $\alpha=1$ ($\beta=0$)
is the case where the first clone is perfect, whereas $\beta=1$
($\alpha=0$) is the case where the second clone is perfect.
The normalization relation implies that
\beq
|\alpha|^2+{2\over N} {\rm Re}(\alpha\beta^*)+|\beta|^2=1
\eeq
which characterizes the balance between the quality of the two clones.
In particular, the {\em symmetric} universal $N$-dimensional cloner
corresponds to the case where
\beq
\alpha^2=\beta^2={N \over 2(1+N)}
\eeq
Using Eq.~(\ref{F_A}) for the cloning fidelity in the computational basis
(since all states are copies with the same fidelity),
we recover the standard formula
for the universal cloners\cite{buzek98,werner,cerf-acta,cerf-jmo}
\beq
F=\left(\alpha+{\beta\over N}\right)^2+(N-1)\left({\beta\over N}\right)^2
={3+N \over 2(1+N)}
\eeq
In particular, the symmetric universal qutrit cloner ($N=3$)
is characterized by a fidelity of 3/4.

\section{Conclusion}

We have investigated several categories of $1\to 2$ cloning transformations
for a three-dimensional system (a qutrit). First, we have analyzed
the cloners that optimally copy the states of any two out of the four
maximally-conjugate bases. The symmetric cloner of this class
has a cloning fidelity of ${1\over 2}+{1\over\sqrt{12}}\simeq 0.789$.
Second, we studied the cloners that copy equally well and with the highest
fidelity three maximally-conjugate bases. 
Actually, these cloners can be shown to copy all states of the
form ${1\over\sqrt 3} (\ket{0}+e^{i\alpha}\ket{1}+e^{i\beta}\ket{2})$
with the same fidelity for any $\alpha$ and $\beta$,
so they are the natural extension of the phase-covariant
qubit cloners. The symmetric cloner of this class copies all these states
with a fidelity ${5+\sqrt{17}\over 12} \simeq 0.760$, and coincides with
the so-called double-phase covariant qutrit cloner analyzed independently
in Ref.~\cite{presti}.
Finally, the cloners that optimally copy all four maximally-conjugate bases
can be shown to copy all states of a qutrit equally well, so they simply
correspond to the universal qutrit cloners. 
The symmetric universal qutrit cloner
has a fidelity of 3/4, in accordance 
to Ref.~\cite{buzek98,werner}. 
We conclude thus that, quite naturally, the cloning fidelity decreases 
when we put a stronger requirement
on the cloner (namely two, three, or four bases must be copied optimally).
This study also suggests that there is still much room for further 
investigation on multi-level non-universal quantum cloning.

\bigskip
\bigskip
\leftline{\Large \bf Appendices}

\appendix
\section{Phase-covariant cloner for a qubit}
\label {qubit}

In this Appendix, we show that the phase-covariant qubit cloner~\cite{bruss}
can be obtained in just a few lines by using the general characterization
of Pauli cloners of Refs.~\cite{cerf-acta,cerf-jmo,cerf-prl,cerf-nasa}.
The phase-covariant qubit
cloner is defined as a transformation that optimally copies all states of the
form ${1\over \sqrt{2}}(\ket{0}+e^{i\alpha}\ket{1})$ for any $\alpha$.
Here, we rather look for a qubit cloner that copies
any two maximally-conjugate bases.
Actually, in the Hilbert space of a qubit, there are three maximally-conjugate
bases, which correspond to the eigenstates of the three Pauli matrices:
\beqa
&&\ket{0} ,\qquad \ket{1}\\
&&\ket{0'}={1\over\sqrt{2}}(\ket{0}+\ket{1}),\qquad
\ket{1'}={1\over\sqrt{2}}(\ket{0}-\ket{1}) \\
&&\ket{0''}={1\over\sqrt{2}}(\ket{0}+i\ket{1}),\qquad
\ket{1''}={1\over\sqrt{2}}(i\ket{0}+\ket{1})
\eeqa
The universal qubit cloning machine \cite{buzek} copies
the states of each of these three bases with the same fidelity.
In contrast, the cloner we will be interested in here is required
to optimall copy only the first two bases with the same (and maximum)
fidelity. This is equivalent to requiring that the states
that on the ``equatorial'' plane $x$-$z$ of the Bloch sphere
are all shrunk by a same factor. (Note that, conventionally, the
phase-covariant
qubit cloner is rather required to optimally copy the last two bases,
or, by extension, all states of the equatorial plane $x$-$y$ \cite{bruss}.)
The cloning fidelity can be higher than that of the universal cloner,
but this is at the expense of cloning fidelity for the third basis,
wich must be lower.

Let us calculate the fidelity of this phase-covariant cloner. Let us
consider the effect of the error operators on the elements of the
two first bases. We have
\beq
\begin{array}{ll}
U_{0,0} \begin{array}{l} \ket{0}\to\ket{0} \\ \ket{1}\to\ket{1} \end{array}
&
~~~
U_{0,1} \begin{array}{l} \ket{0}\to\ket{0} \\ \ket{1}\to-\ket{1} \end{array}
\\[0.5cm]
U_{1,0} \begin{array}{l} \ket{0}\to\ket{1} \\ \ket{1}\to\ket{0} \end{array}
&
~~~
U_{1,1} \begin{array}{l} \ket{0}\to\ket{1} \\ \ket{1}\to-\ket{0} \end{array}
\end{array}
\eeq
so the elements of the first basis are left unchanged (up to
a sign) by the error operators $U_{0,0}$ and $U_{0,1}$.
Similarly, we have
\beq
\begin{array}{ll}
U_{0,0} \begin{array}{l} \ket{0'}\to\ket{0'} \\ \ket{1'}\to\ket{1'} \end{array}
&
~~~
U_{0,1} \begin{array}{l} \ket{0'}\to\ket{1'} \\ \ket{1'}\to\ket{0'} \end{array}
\\[0.5cm]
U_{1,0} \begin{array}{l} \ket{0'}\to\ket{0'} \\ \ket{1'}\to-\ket{1'}\end{array}
&
~~~
U_{1,1} \begin{array}{l} \ket{0'}\to -\ket{1'} \\ \ket{1'}\to \ket{0'}
\end{array}
\end{array}
\eeq
so the elements of the second basis are left unchanged (up to a sign)
under $U_{0,0}$ and $U_{1,0}$.
Now, using the general formula for the cloning fidelity Eq.~(\ref{fidelity}),
we find that the elements $\ket{0}$ and $\ket{1}$ of the first basis
are cloned with the fidelity
\beq
F=p_{0,0}+p_{0,1}
\eeq
while the elements $\ket{0'}$ and $\ket{1'}$ of the second basis
are cloned with the fidelity
\beq
F'=p_{0,0}+p_{1,0}
\eeq
The requirement of having a phase-covariant cloner ($F=F'$)
can thus be simply written as $p_{0,1}=p_{1,0}$. Consequently, we simply
consider a cloner characterized by the amplitude matrix
\beq
(a_{m,n})=\left(
\begin{array}{ll}
v & x \\
x & y
\end{array}
\right)
\eeq
where $x$, $y$ and $v$ are real and positive, and with the
normalization condition $v^2+2 x^2+y^2=1$. The fidelity $F$
(and disturbance $D=1-F$) of the first clone
are thus given in both bases by
\beqa
F&=&F'=v^2+x^2 \\
D&=&D'=x^2+y^2
\eeqa
For the second clone, Eq.~(\ref{FT}) [or, equivalently, Eq.~(\ref{TR2})],
implies that the matrix $b_{m,n}$  has the same form as
$a_{m,n}$ with the
substitution
\beqa
v &\to& (v+2x+y)/2 \\
x &\to& (v-y)/2 \\
y &\to& (v-2x+y)/2
\eeqa
so that the states of the two conjugate bases are again copied all with
a same fidelity (and a same disturbance):
\beqa
\tilde F &=& (v^2+2x^2+y^2+2vx+2xy)/2 = 1/2 + vx+xy \\
\tilde D &=& (v^2+2x^2+y^2-2vx-2xy)/2 = 1/2 - vx-xy
\eeqa
We are now interested in finding the cloner that maximizes the fidelity
of the second clone ${\tilde F}$ for a given fidelity of the first clone $F$.
A simple constrained maximization calculation yields the solution
\beqa
v &=& F \\
x &=& \sqrt{F(1-F)} \\
y &=& 1-F
\eeqa
so that the maximum fidelity of the second clone can be written as
a function of the fidelity of the first clone
\beq
{\tilde F}= {1\over 2} + \sqrt{F(1-F)}
\eeq
This expresses the balance between the quality of the two clones
in the case of a phase-covariant qubit cloner.
As expected, $F=1$ yields ${\tilde F}=1/2$, and conversely. The symmetric
phase-covariant cloner yields two clones of equal fidelity
\beq
F={\tilde F}={1\over 2}+{1\over\sqrt{8}} \simeq 0.854
\eeq
in agreement with \cite{bruss}. As expected, this fidelity is slightly
higher than
the fidelity of the universal qubit cloner, namely F=5/6. In contrast, the
third
basis is now copied with a fidelity equal to 3/4, that is, lower than the
fidelity of the universal cloner.

\bigskip
\leftline{\Large \bf Acknowledgment}

We are grateful to G. M. D'Ariano and P. Lo Presti for communicating us
their preliminary results on the double-phase covariant qutrit cloner.
T.D. is a Postdoctoral Fellow of the Fonds voor Wetenschappelijke Onderzoek,
Vlaanderen. N.C. acknowledges funding by the European Union under the
project EQUIP (IST-FET programme).

\end{document}